\documentclass{PoS-hep}
\usepackage{amsmath}
\usepackage{bm}
\usepackage{epsfig}
\usepackage{graphics}
\usepackage{eufrak}
\usepackage{cite}
\usepackage{epsfig}
\usepackage{amssymb}
\usepackage{latexsym}
\usepackage{slashed}
\usepackage{upgreek}
\usepackage{amsmath}
\usepackage{amsfonts}
\usepackage{amsbsy}
\usepackage{amscd}
\usepackage{bbm}

\def\lt{\textgreek}

\usepackage[english,polutonikogreek]{babel}
\usepackage[iso-8859-7]{inputenc}

\newcommand{\eec}{\end{center}}
\newcommand{\bec}{\begin{center}}

\newcommand{\eem}{\end{matrix}}
\newcommand{\bem}{\begin{matrix}}
\newcommand{\eeq}{\end{equation}}
\newcommand{\beq}{\begin{equation}}
\newcommand{\ba}{\begin{array}}
\newcommand{\ea}{\end{array}}
\newcommand{\bea}{\begin{eqnarray}}
\newcommand{\eea}{\end{eqnarray}}
\newcommand{\baq}{\begin{eqnarray}}
\newcommand{\eaq}{\end{eqnarray}}
\newcommand{\beqs}{\begin{subequations}}
\newcommand{\eeqs}{\end{subequations}}
\newcommand{\bel}{\begin{align}}
\newcommand{\eal}{\end{align}}

\newcommand{\ftn}{\footnotesize}

\newcommand{\etal}{{\it et al.\/}}

\def\to{\rightarrow}

\newcommand{\Glr}{\ensuremath{\mathbb{G}_{\rm LR}}}

\newcommand{\Gsm}{\ensuremath{\mathbb{G}_{\rm SM}}}
\newcommand{\Gps}{\ensuremath{\mathbb{G}_{\rm PS}}}

\newcommand{\sur}{\ensuremath{SU(2)_{\rm R}}}

\def\ssb{\leavevmode\hbox{$\diagup$\kern-12pt\ftn\scshape
susy}}

\newcommand{\hepph}[1]{{\small\tt hep-ph/#1}}

\newcommand{\arxiv}[1]{{\small\tt  arXiv:#1}}

\newcommand{\fref}[1]{Fig.~\ref{#1}}

\newcommand{\Sref}[1]{Sec.~\ref{#1}}
\newcommand{\Fref}[1]{Fig.~\ref{#1}}

\newcommand{\cref}[1]{Ref.~\cite{#1}}

\newcommand{\bdhh}{{\ensuremath{\normalsize I{\kern-2.9pt H}}}}

\newcommand{\phc}{\ensuremath{\Phi}}
\newcommand{\phcb}{\ensuremath{\bar\Phi}}

\newcommand{\hh}{{\ensuremath{
I{\kern-2.6pt h}}}}
\newcommand{\bhh}{{\ensuremath{\bar{
I{\kern-2.6pt h}}}}}

\newcommand{\tnb}{{\ensuremath{\tan\beta}}}

\renewenvironment{subequations}{%
\refstepcounter{equation}%
\setcounter{parentequation}{\value{equation}}%
  \setcounter{equation}{0}
  \ignorespaces
}{%
  \setcounter{equation}{\value{parentequation}}%
  \ignorespacesafterend
}

\title{\boldmath \bfseries Memories of Prof. George Lazarides}

\author{\speaker{C. Pallis}\\
School of Technology, \\
Aristotle University of Thessaloniki, \\ GR-541 24 Thessaloniki, GREECE \\
E-mail: \email{kpallis@auth.gr}}

\abstract{I present some memories of my Ph.D supervisor and,
later, collaborator but always encouraging supporter Prof. G.
Lazarides. Some of his contributions to our common and related
scientific activities on the phenomenology of MSSM and inflation
are also summarized.\\

\centering{\includegraphics[height=1.3in,angle=-90]{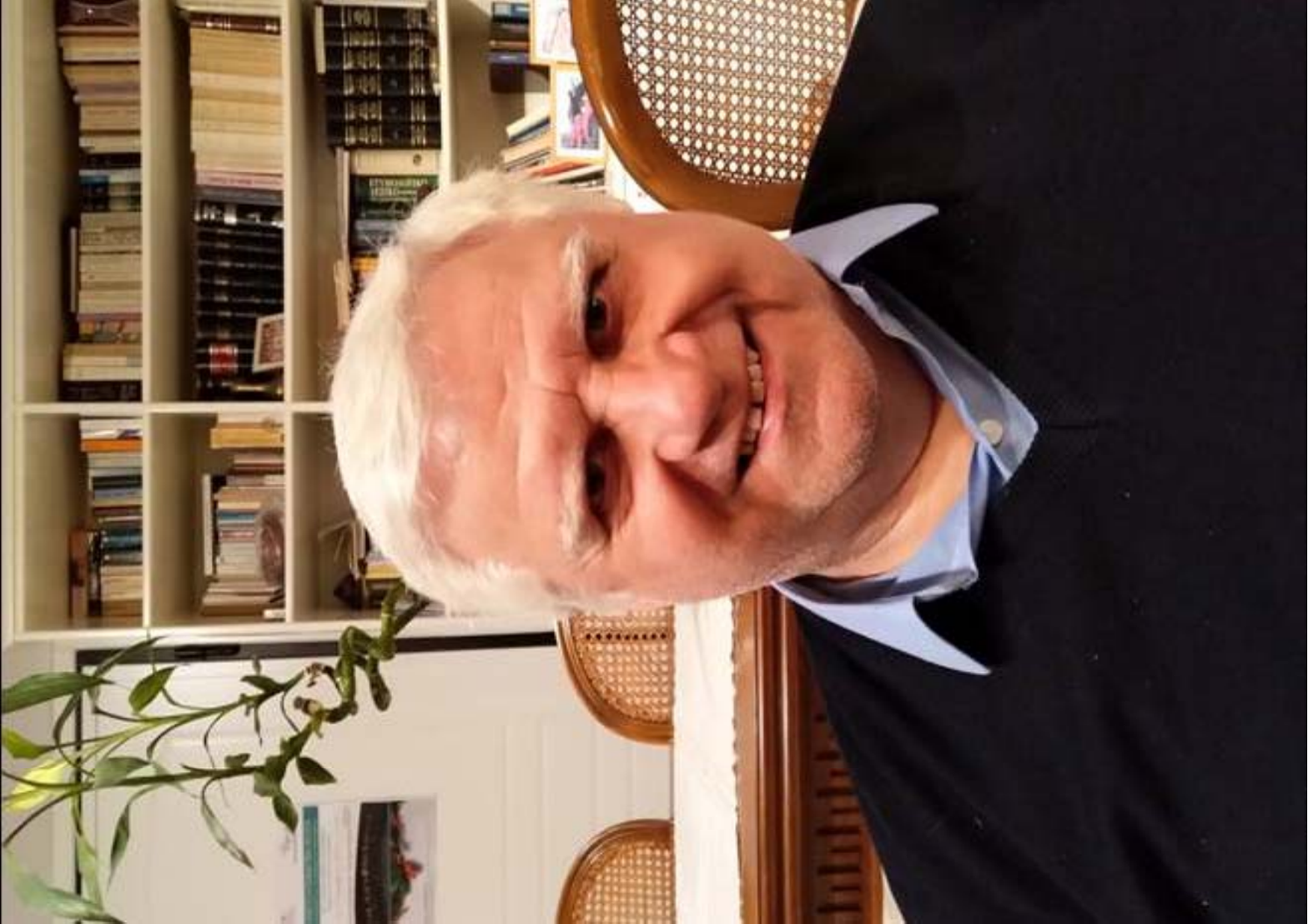}}
\\[.8cm] {\sl\bfseries Published in}~~{PoS  CORFU {\bf
2025}, 154 (2026)}. }

\FullConference{Corfu Summer Institute 2025 "School and Workshops
on Elementary Particle Physics and Gravity" (CORFU2025)
24 August - 3 September, 2025\\
Corfu, Greece}

\ShortTitle{Memories of Prof. G. Lazarides}

\begin{document}

\selectlanguage{english}

It is a pleasure and an honor for me to share my memories of Prof.
G. Lazarides, with whom I was associated by a (periodically very
close) ``studentship'' of about thirty years. Here, I attempt a
retrospect trying to express my admiration and gratitude. For
organizational purposes, I divide the text into three sections
(Sec.~\ref{sec1} -- \ref{sec3}), corresponding to the periods of
our interaction, and let as an epilogue in \Sref{ep} some thoughts
on his multi-dimensional personality.


\section{Ph.D Period}\label{sec1}

I met Prof. Lazarides early in 1994 as a student in the Physics
Department of the \emph{Aristotle University of Thessaloniki}
({\sf\small AUTH}) after a recommendation of Prof. N.D. Vlachos.
He was the representative of the Erasmus Program in the School of
Technology at AUTH. In the framework of that program I performed a
three-month visit at the ``\'Ecole Polytechnique'' in Paris aiming
to prepare a diploma thesis under the supervision of Prof. K.
Bachas. From our first meeting I remember his alert: ``You should
work in Paris!'' After years, when I read his recommendation
letter I relieved by his remark: ``Costas is an exceptionally hard
worker.''

\begin{floatingfigure}[r]\vspace{-2.5cm}
\centering{\includegraphics[height=3.2in]{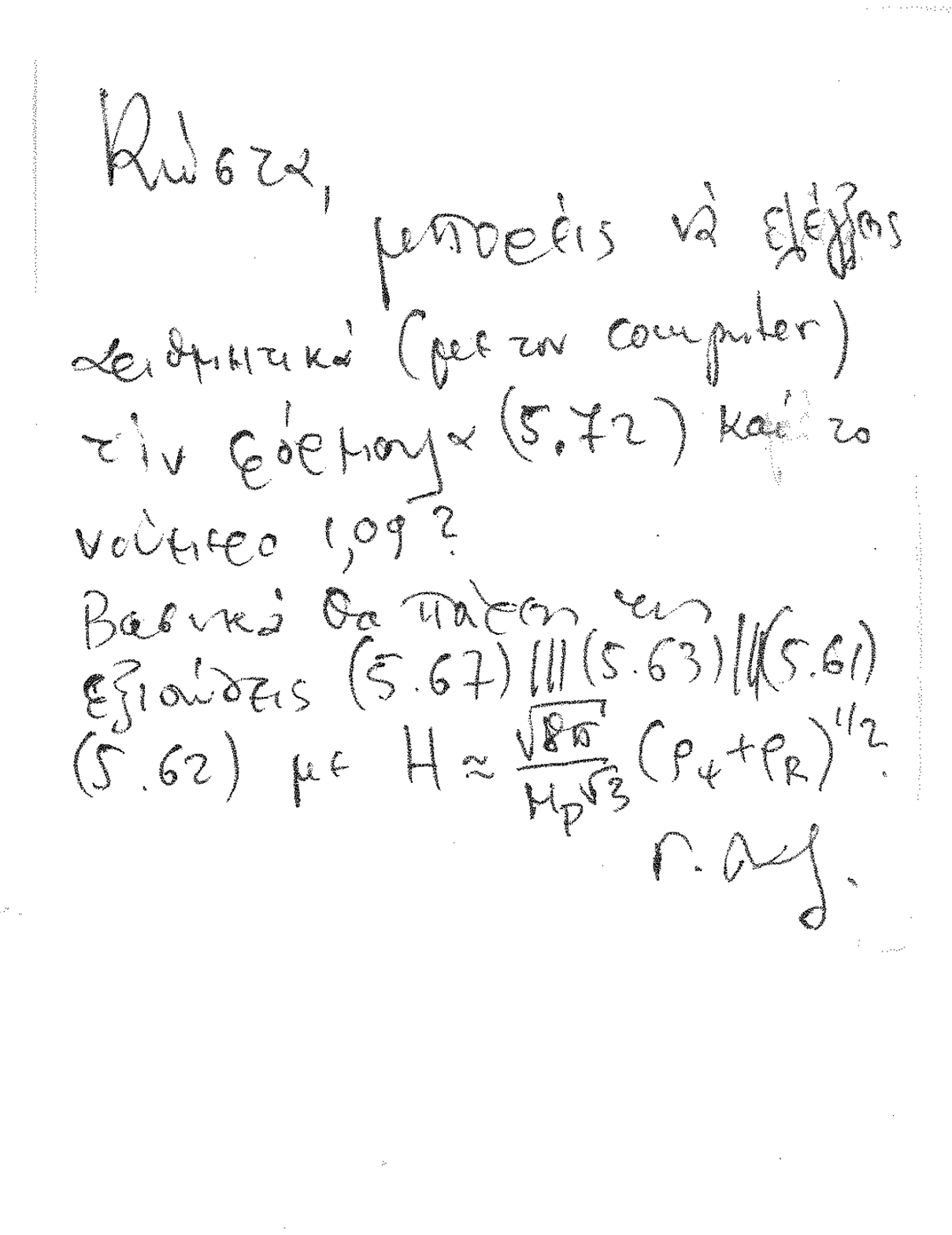}}
\caption {\sl\small A handmade message asking me to verify
numerically a factor in Eq. (5.72) of Ref.~[1].}\label{note}
\end{floatingfigure}

After about a year, he accepted me to work on a doctoral thesis
with him. ``Costas is my student'' he was saying with confident.
The topic he chose was extremely interesting and timely. In
particular, motivated by one of his seminal paper \cite{anant} and
a companion one \cite{anant1}, he was planning to investigate the
problem of \emph{Cold Dark Matter} ({\sf\small CDM}) within the
\emph{Constrained Minimal Supersymmetric ({\sf\small SUSY})
Standard Model} ({\sf\small CMSSM}) endowed with the \emph{Yukawa
(coupling) unification} ({\sf\small YU}). I started working on
this project alone around June and from October with the
collaboration of Mario G\'omez who joined us as a postdoctoral
researcher. Around February and then in May we came across with
two papers from J. Ellis \etal\ \cite{ellis, ellis1} which analyze
the relic density of the neutralino \emph{Lightest SUSY Particle}
({\sf\small LSP}) within CMSSM in the low $\tan\beta$ regime
taking into account not only annihilation processes between
neutralinos but also coannihilation ones between neutralinos and
the little heavier sleptons. We included these extra contributions
in our analysis for large $\tnb$ values and completed in June our
paper which became quickly well-recognized. The citations we
received from it accompany me during my trajectory. I remember
Prof. Lazarides with his little calculator to derive the relative
contributions of the various processes to the LSP relic density
and insist on avoiding any ``discrepancy'' between our results.

Combining the CDM considerations with phenomenological
restrictions arising from the SUSY corrections to $b$-quark mass
and the $b\to s\gamma$ branching ratio we succeeded in our second
paper \cite{cd2} to revitalize temporarily CMSSM with YU taking
the correct sign of the $\mu$ parameter. Finally, a third
publication \cite{cd3} was realized by applying the total system
of constraints in a version of CMSSM with initial conditions for
the soft SUSY breaking parameters from the Ho\v rawa-Witten
theory.

At the final stage of my thesis \cite{phd} Prof. Lazarides
empathized with me the stress to complete and defend it before go
for my military service -- note that Greek laws were very strict
regarding this date. The unique time that I listened to him
emotionally moved was when he walked me to the door of his office
after the obtention of my Ph.D degree. I still hear his words:
``Ante re -- untranslated from Greek --, become a good soldier
now!''

\section{Post-Doctoral Period}\label{sec2}

During my military service it was a pleasure for me to visit him
when I had leave of absence. We initially revised our last paper
\cite{cd3} and then we pursued our collaboration on YU within
CMSSM. I remember him in a phone call from the army station in
Kozani to announce me that he found a way to violate moderately YU
so that we simultaneously satisfy all the existing
cosmo-phenomenological restrictions.

\begin{figure}[!t]\vspace*{-.17in}
\hspace*{-.12in}
\begin{minipage}{8in}
\includegraphics[height=3.2in,angle=-90]{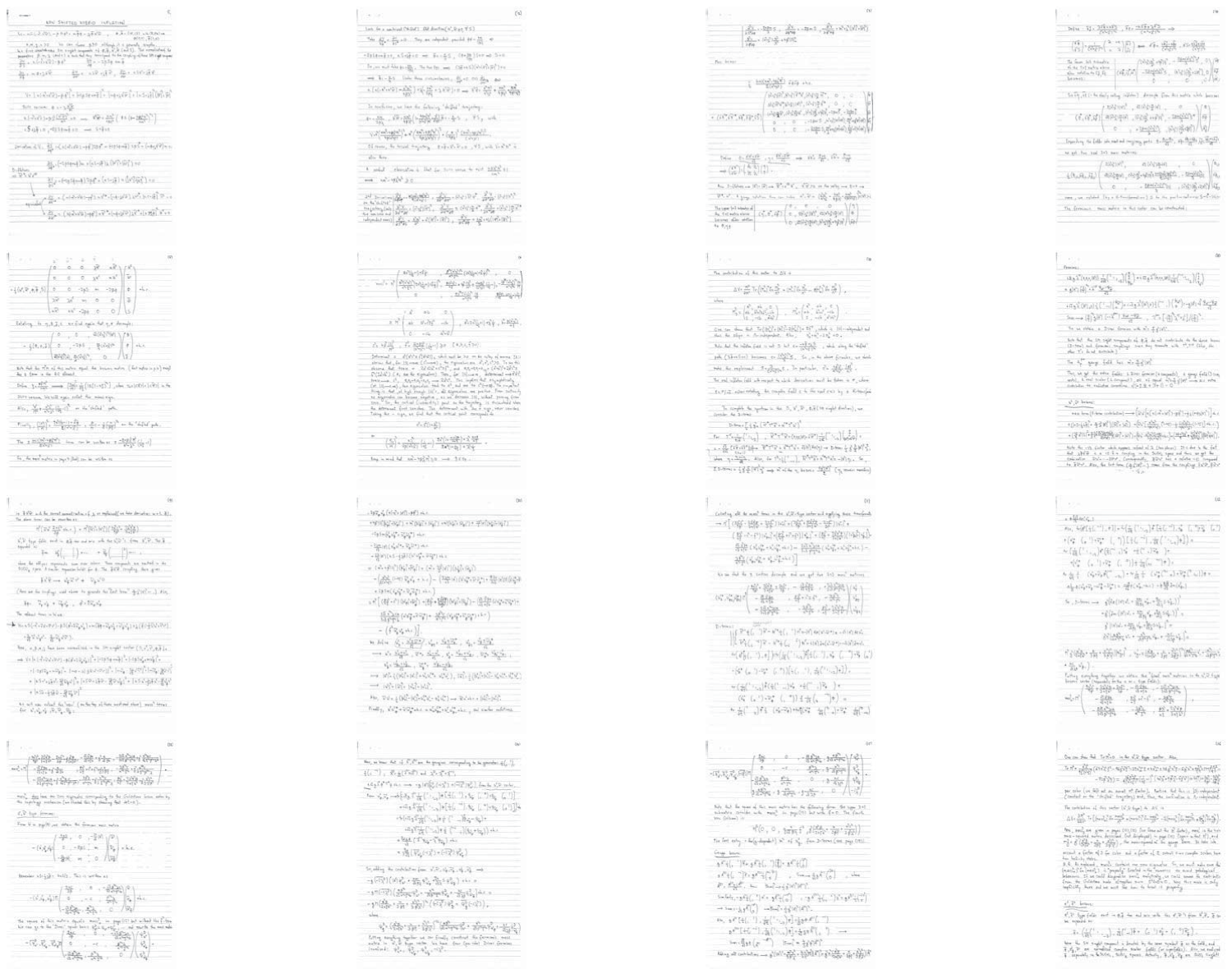}
\hspace*{-.2cm}
\includegraphics[height=3.2in,angle=-90]{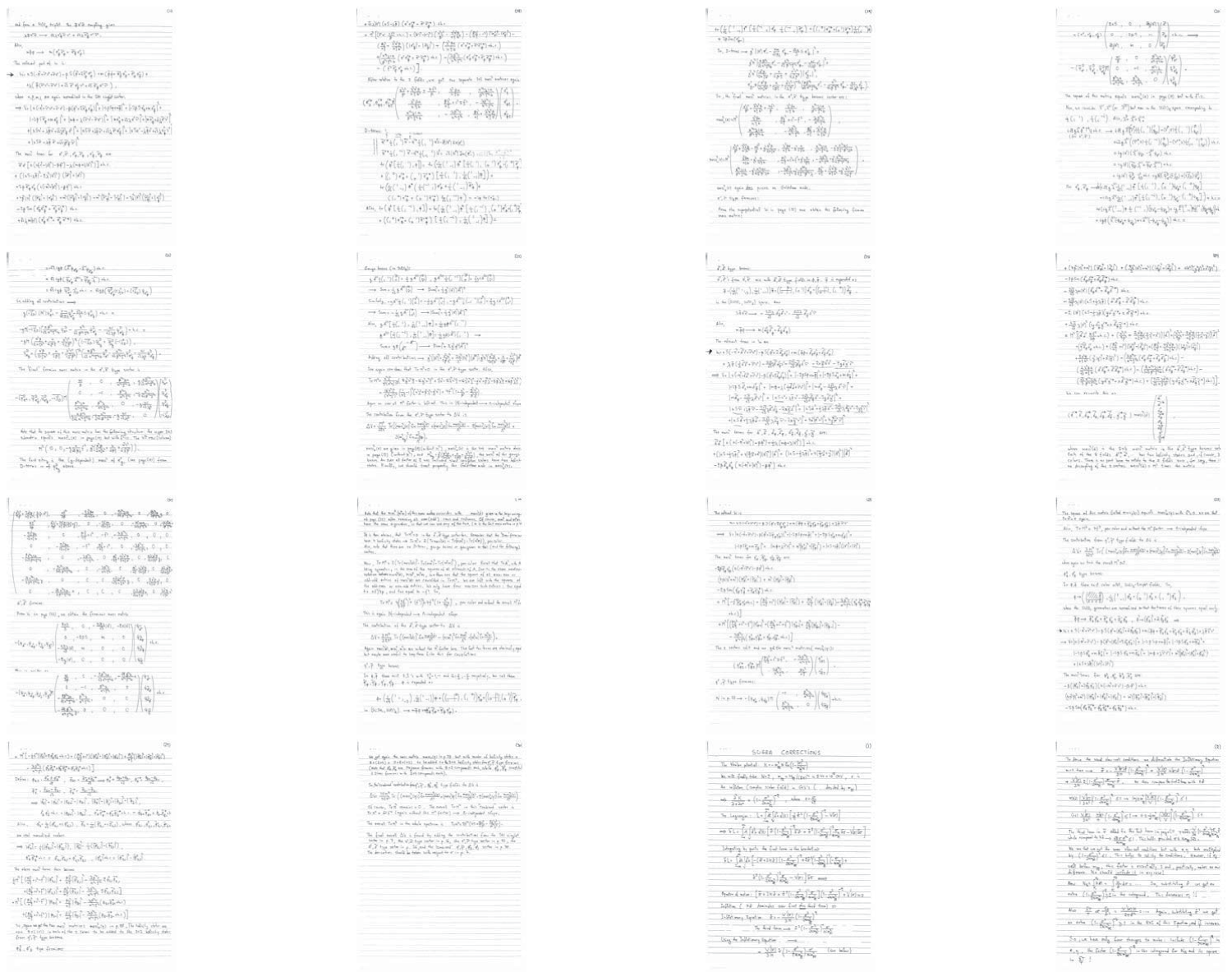}
\end{minipage}
\hfill\caption{\sl Radiative corrections -- derived by hand (!) -
during new-shifted FHI \cite{nsh}.}\label{fig1}
\end{figure}

Indeed, working in the context of Pati-Salam gauge group $\Gps$,
he extended the simplest version of the model \cite{leo}, which
predicts YU since the trilinear terms of the superpotential of
MSSM, $W_{\rm MSSM}$, originate from a unique coupling between the
matter fields $F_i$ and $F^c_i$ and the electroweak Higgs
bidoublet $\hh$. He included two new bidoublets $\hh'$ and $\bhh'$
in the $({\bf 15, 2, 2})$ representation -- from which $\hh'$ may
couple to $F_i$ and $F^c_i$ -- and two $\sur$ triplets $({\bf 15,
1, 3})$ $\phc$ and $\phcb$ and/or two $\sur$ singlets $({\bf 15,
1, 1})$ $\phc'$ and $\phcb'$. During the $\Gps$ breaking $\phc$
and $\phc'$ acquire superheavy \emph{vacuum expectation values}
({\sf\small v.e.vs}) and generate a mixing between the doublets in
$\hh$ and $\hh'$ thanks to their couplings with them. So, we
obtain one pair of superheavy doublets with zero v.e.vs and one
pair which remains massless at the GUT scale and can be identified
with the electroweak doublets. As a consequence, the trilinear
terms of $W_{\rm MSSM}$ now originate from two terms including
$\hh$ and $\hh'$. Therefore, exact YU is naturally replaced by a
number of \emph{Yukawa quasi-unification} ({\sf\small YQU})
conditions which allow for lower but rather large $\tnb$ values.
The cosmo-phenomenology of CMSSM endowed with YQU is analyzed in
\cref{quasi,quasi1,nick1} before the discover of Higgs boson and
after it in a generalized framework \cite{nick2,nick3,nick4}. The
work of the four last papers was performed with the collaboration
of N. Karagiannakis, who completed his Ph.D thesis \cite{phdn}
under the supervision of Prof. Lazarides.

However, this was not the end of the story! The part of the
superpotential which is relevant for the breaking of $\Gps$ to
$\Gsm$ may support a cornucopia of inflationary scenarios. Namely,

\begin{itemize}


\item New -- in contrast to the original \cite{sh} -- shifted
\emph{F-term hybrid inflation} ({\small\sf FHI}) \cite{nsh} where
$\Gps$ is totally broken down to $\Gsm$. Due to the large
representations of $\Gps$ the computation of the relevant particle
mass spectrum, needed for the estimation of the radiative
corrections to the inflationary potential, is really involved.
Lazarides carried out this task by hand -- see notes in
\Fref{fig1} -- checking explicitly the independence of the result
from the renormalization scale.

\item New -- in contrast to the original \cite{sm, sm1} --  smooth
FHI \cite{nsm} based only on renormalizable terms where again
$\Gps$ is totally broken down to $\Gsm$.

\item Semi-Shifted FHI \cite{ssh}, where $\Gps$ is subsequently
broken down to Left-Right gauge group $\Glr$ during FHI  and then
to $\Gsm$ producing cosmic strings.

\item Standard-smooth FHI \cite{stsm}, where standard FHI
\cite{fhi} is followed by a stage of new smooth FHI assisting in
obtaining the correct value of the scalar spectral index -- see
\cref{com}. The work of the above papers was performed with the
collaboration of A. Vamvasakis, who completed his Ph.D thesis
\cite{phdax} under the supervision of Prof. Lazarides.

\item Standard-semi-shifted FHI \cite{stssh}, where standard FHI
\cite{fhi} is followed by a stage of semi-shifted FHI. As a result
we obtain monopoles, partially inflated away, and cosmic strings
which may become quasi-stable leading to gravitational waves.

\end{itemize}

Similar superpotential terms have been also considered in the
context of $\Glr$ \cite{lrtfhi, lrthi}. His overall conclusion of
these works was that inflation is yet another motivation for SUSY,
thanks to the plethora of flat directions which arise in the
moduli space and facilitate its establishment.

\begin{figure}[!t]\vspace*{-.15in}
\centering{\includegraphics[height=1.8in]{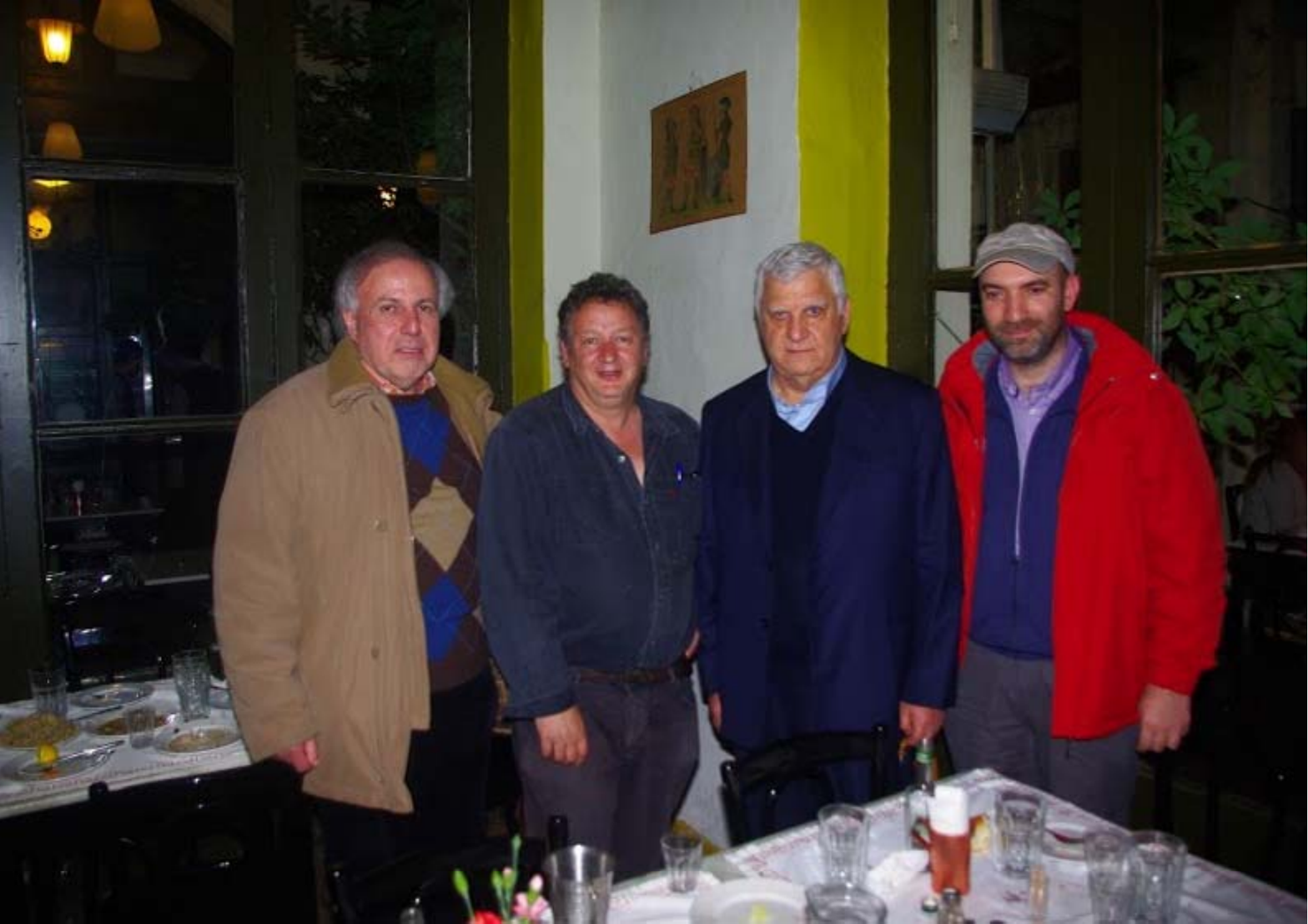}}
\caption[]{\sl\small With C. Panagiotakopoulos, M.E. G\'omez and
me during the George-Fest on March 2013.}\label{fig2}
\end{figure}

During my postdoctoral adventures abroad, Prof. Lazarides was a
person of reference.  His countless answers to questions -- see
e.g. the left panel of \fref{fig4} --, not always related to our
joint publications, and his support in my difficulties, not only
academic, were invaluable. In each question he had the courage of
his opinion and clear research orientations. I always returned to
him asking inspiration, scientific guidance and encouragement in
life difficulties. He always welcomed me in his office ready to
become a competitor in research and a companion in difficult
times. I was leaving his office wiser and more prudent.


\newpage

\section{HFRI Period}\label{sec3}

Our last cooperation was in a research program of the
\emph{Hellenic Foundation of Research and Innovation} ({\small\sf
HFRI}). Thanks to his rich CV, our research proposal was ranked
second among sixteen funded in the Scientific Field ``Natural
Sciences'' and hundreds submitted. He worked hard within the
framework of the program, although unpaid, and leaves us a legacy
of many works which -- coincidentally? -- completed topics he had
delved into as a young researcher \cite{magg, kibble, kibbledw}.
Namely, he focused with the collaboration of Q. Shafi and other
younger researchers on the generation of cosmological defects
\cite{so,sotw} -- such as cosmic strings, monopoles, domain walls
and combinations of them -- within \emph{Grand Unified Theories}.
Some of these objects emit during their cosmological evolution
gravitational waves which may interpret the recent \emph{pulsar
timing array} results \cite{nano}. Although, I did not participate
in those publications -- due to the pandemic isolation and my
teaching duties --, I was in regular communication with him to
accomplish the management of the program.

\begin{figure}[!t]
\begin{tabular}[!h]{cc}\begin{minipage}[t]{7.5cm}
\vspace*{-2.4in}\centering{\includegraphics[height=4.8in]{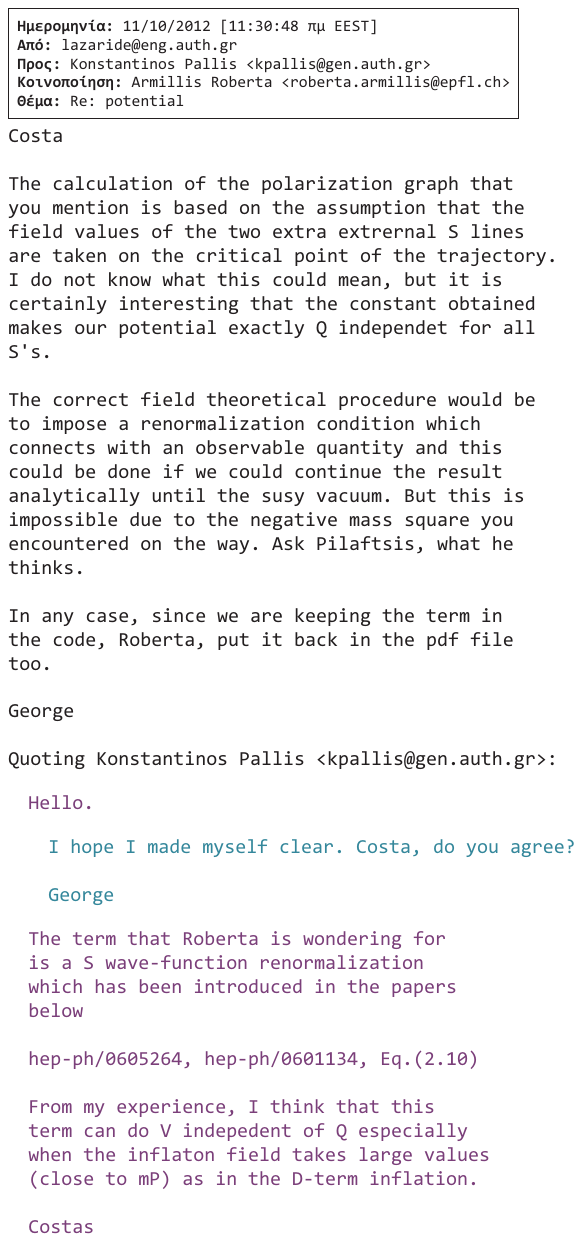}}\end{minipage}
&\begin{minipage}[h]{7.5cm} 
\centering{\includegraphics[height=3.2in]{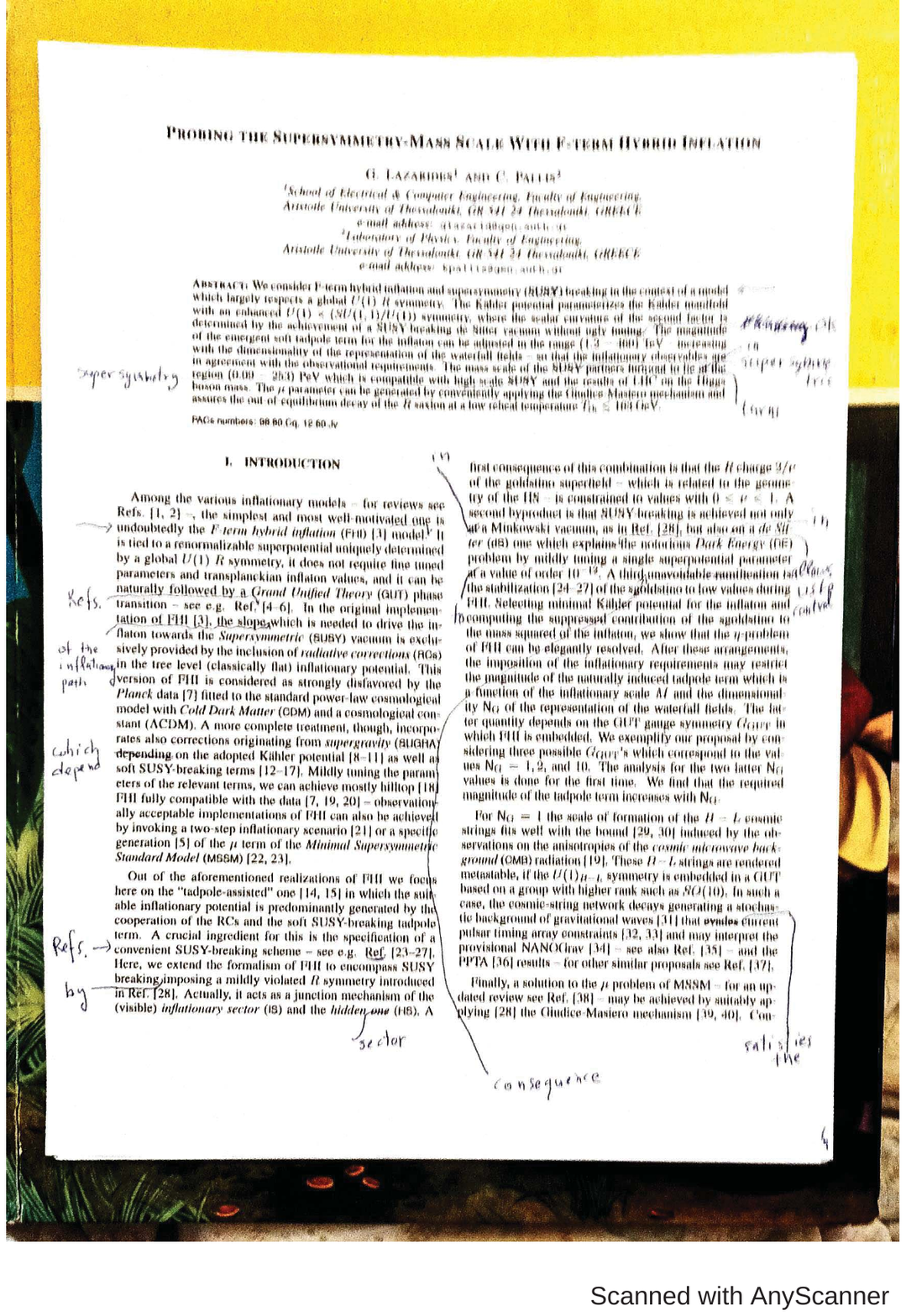}}\end{minipage}
\end{tabular} \vspace*{-1.5cm}
\caption[]{\sl\small Message exchange related to \cref{lrtfhi}
(left panel) and corrections on the manuscript of our last paper
\cite{asfhi} (right panel).}\label{fig4}
\end{figure}

It was in this reexamining and perhaps nostalgic moment that he
was struck down by illness. Even when he learned of his health
situation, he did not propose someone else as Scientific
Coordinator for the program, since he did not want to disrupt the
conditions under which the members of the research team
participated. Moreover, he continued to work, despite his
sickness, being in the cutting edge of his research. Divine
Providence allowed us to complete one last work \cite{asfhi} --
some of his corrections are shown in the right panel of
\fref{fig4} -- and submit the final report of the program on time.
In his last message on ninth February 2024 he urged me to ask
citations for our paper. He was worried about my survival in the
field of research...


\newpage

\section{Epilogue} \label{ep}

Let me close this presentation by outlining some salient aspects
of the charismatic and multifaceted personality of G. Lazarides as
a physicist, in \Sref{phy}, and beyond Physics in \Sref{bey}.

\subsection{His Principles as Physicist}\label{phy}

Prof. Lazarides was actually the incarnation of the word
``professor'' taking into account the original etymology of the
corresponding greek word ``\lt{καθηγητής}'' (kathigitis in
greeklish). It comes from the verb ``\lt{<ηγο\~υμαι}'' (igoume in
greeklish) associated with the preposition ``\lt{kaτά}'' and means
``I lead and guide''. Indeed, he was an inspiring world-class
leader and a paradigmatic guide/educator. Most importantly, he was
instilling with his attitude honesty, expressional accuracy and
sincerity in research. In our collaborations he was justifiably
demanding, exhaustively perfectionist and inexhaustibly dynamic
when he defended our common work. Needless to say, he was
admitting his possible mistakes and he was rectifying them with
constructive disposition. The final output was satisfactory for
all participants and it was often being recognized by the
international community with numerous citations. Despite his
prominent scientific achievements, he remained a low-profile man,
away from early enthusiasms and unnecessary voices. He knew that
scientific research is constantly under revision and development.


Although he had a very strong mathematical background and he
wanted to be mathematically consistent in his works, he was
aspiring to concentrate on cosmo-phenomenological problems of the
``real world'' and provide clear solutions to them without being a
computationally precision lover. He believed that physics should
be connected to experiments and observations. He was following the
mean stream in the science giving priority to the serious
motivation of a work. Also, he was preferring to work in the one
side of a dipole than to participate in both sides of it. E.g., in
2018 he told me that he prefers better high-scale SUSY than not at
all SUSY. Finally, he was relatively conservative in the issues he
decided to deal with. E.g., his favor inflationary model was FHI
\cite{fhi} and its variants \cite{pqfhi,asfhi,lrtfhi} which are
based on the SUSY framework and require just the inclusion of some
supergravity corrections. Indeed, despite he advised me to improve
my paper \cite{r2} on Starobinsky inflation and he participated in
a paper \cite{hi} on Higgs inflation, he was sceptical with these
models since they entail somehow radical (super-)gravity
manipulations.

\subsection{Beyond Physics}\label{bey}

In addition to being an outstanding physicist, George Lazarides
had many other interests.  He was a scholar of Byzantine history
and art and a frequent visitor of Mount Athos with theological
culture and concerns. He was, also, a lover of the Greek language
and an excellent user of it. On the other hand, he was a theorist
of rebetiko (a stream of Greek popular music in the decade of
1950) and an enthusiastic learner of bouzouki (a musical
instrument used at that times) the last years. Both of his hobbies
were in harmony with the unseparate from his hands komboloi
(worry-beats).

Last but not least, he had a rare general education and a natural
sense of humor. As a consequence, he was a unique companion for an
excursion and/or a meal, as Qaisar Shafi vividly sketches in his
contribution \cite{qaisar}. One of the few opportunities I had to
participate in a
\begin{floatingfigure}[hr]\vspace*{-4.8cm}
\hspace*{1.5cm}\includegraphics[height=2.3in]{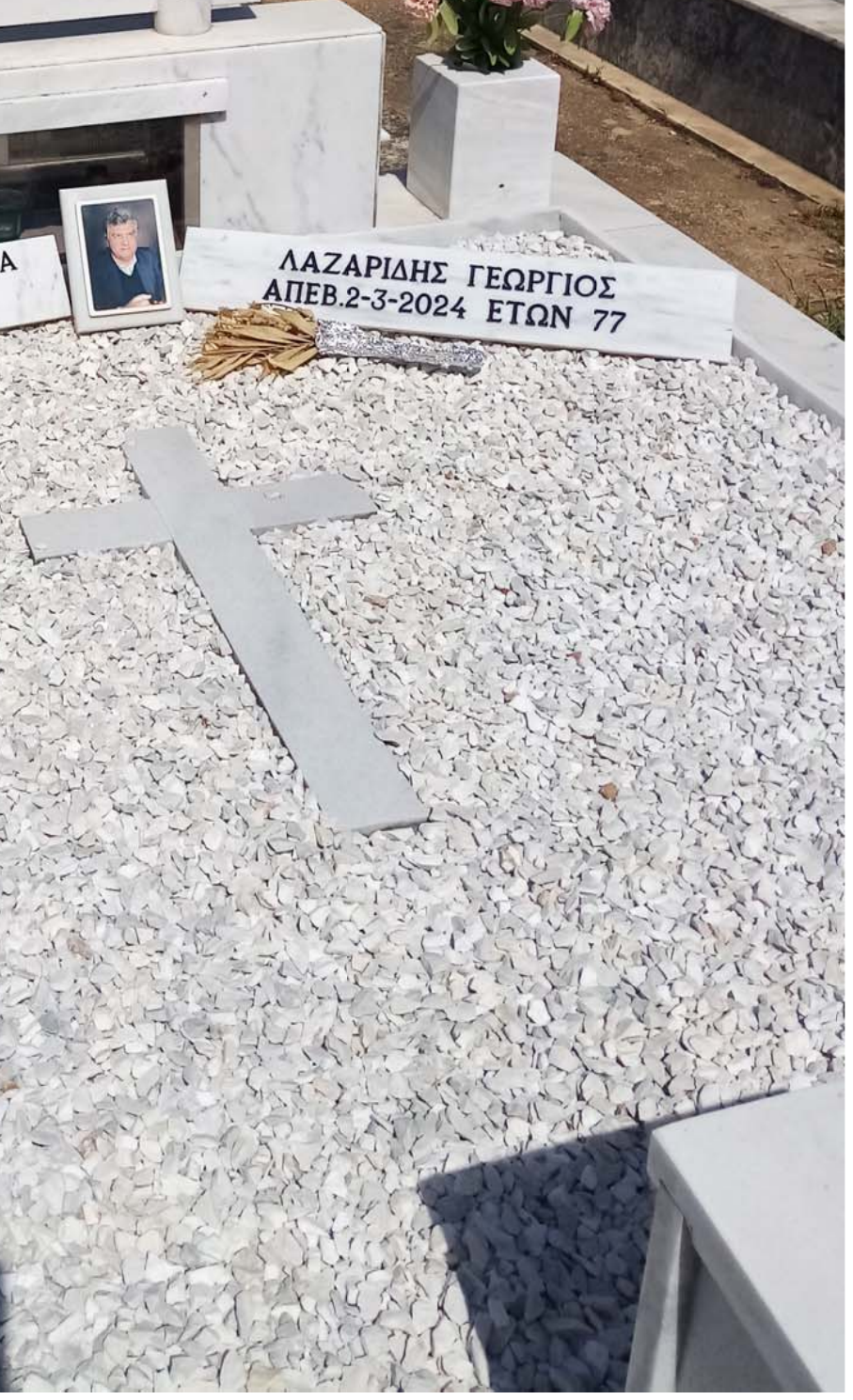}
\\[3.6cm] \caption{\sl\small  May his memory be eternal!}\label{ch}
\end{floatingfigure}
\noindent such unforgettable and treasured event was during his
visit -- together with  Mrs. Elisabeth, his beloved wife whom he
called Veta -- to Cyprus, where I was working as a postdoctoral
researcher. I remember him ``teaching'' mythology in the
archeological area of Paphos and explaining ancient Greek
inscriptions at Kourion and the church of Saint Lazarus in
Larnaca. It was March and he was remembering his father's death...
He said confidentially to me that, when he was kid, since he had
large palms suitable for hand kissing, his father was asking him
fondly: ``How are you, Your Eminence?''

Have a good journey into eternity, Your Eminence...



\def\ijmp#1#2#3{{\sl Int. Jour. Mod. Phys.}
{\bf #1},~#3~(#2)}
\def\plb#1#2#3{{\sl Phys. Lett. B }{\bf #1}, #3 (#2)}
\def\prl#1#2#3{{\sl Phys. Rev. Lett.}
{\bf #1},~#3~(#2)}
\def\rmp#1#2#3{{Rev. Mod. Phys.}
{\bf #1},~#3~(#2)}
\def\prep#1#2#3{{\sl Phys. Rep. }{\bf #1}, #3 (#2)}
\def\prd#1#2#3{{\sl Phys. Rev. D }{\bf #1}, #3 (#2)}
\def\npb#1#2#3{{\sl Nucl. Phys. }{\bf B#1}, #3 (#2)}
\def\npps#1#2#3{{Nucl. Phys. B (Proc. Sup.)}
{\bf #1},~#3~(#2)}
\def\mpl#1#2#3{{Mod. Phys. Lett.}
{\bf #1},~#3~(#2)}
\def\jetp#1#2#3{{JETP Lett. }{\bf #1}, #3 (#2)}
\def\app#1#2#3{{Acta Phys. Polon.}
{\bf #1},~#3~(#2)}
\def\ptp#1#2#3{{Prog. Theor. Phys.}
{\bf #1},~#3~(#2)}
\def\n#1#2#3{{Nature }{\bf #1},~#3~(#2)}
\def\apj#1#2#3{{Astrophys. J.}
{\bf #1},~#3~(#2)}
\def\mnras#1#2#3{{MNRAS }{\bf #1},~#3~(#2)}
\def\grg#1#2#3{{Gen. Rel. Grav.}
{\bf #1},~#3~(#2)}
\def\s#1#2#3{{Science }{\bf #1},~#3~(#2)}
\def\ibid#1#2#3{{\it ibid. }{\bf #1},~#3~(#2)}
\def\cpc#1#2#3{{Comput. Phys. Commun.}
{\bf #1},~#3~(#2)}
\def\astp#1#2#3{{Astropart. Phys.}
{\bf #1},~#3~(#2)}
\def\epjc#1#2#3{{Eur. Phys. J. C}
{\bf #1},~#3~(#2)}
\def\jhep#1#2#3{{\sl J. High Energy Phys.}
{\bf #1}, #3 (#2)}
\newcommand\jcap[3]{{\sl J.\ Cosmol.\ Astropart.\ Phys.\ }{\bf #1}, #3 (#2)}
\newcommand\njp[3]{{\sl New.\ J.\ Phys.\ }{\bf #1}, #3 (#2)}
\def\prdn#1#2#3#4{{\sl Phys. Rev. D }{\bf #1}, no. #4, #3 (#2)}
\def\jcapn#1#2#3#4{{\sl J. Cosmol. Astropart.
Phys. }{\bf #1}, no. #4, #3 (#2)}
\def\epjcn#1#2#3#4{{\sl Eur. Phys. J. C }{\bf #1}, no. #4, #3 (#2)}


\begin{thebibliography}{99}

\bibitem{kolb} E.W. Kolb and M.S. Turner, {\it The Early
Universe}, {\sl Addison-Wesley, Redwood City} (1990).

\bibitem{anant} B.~Ananthanarayan, G.~Lazarides and Q.~Shafi,
{\it Top mass prediction from supersymmetric GUTs,} {\sl Phys.
Rev. D} \textbf{44}, 1613 (1991).

\bibitem{anant1}  B.~Ananthanarayan, Q.~Shafi and X.M.~Wang,
{\it Improved predictions for top quark, lightest supersymmetric
particle, and Higgs scalar masses,} {\sl Phys. Rev. D}
\textbf{50}, 5980 (1994) [\hepph{9311225}].


\bibitem{ellis} J.R.~Ellis, T.~Falk and K.A.~Olive, {\it Neutralino - Stau
coannihilation and the cosmological upper limit on the mass of the
lightest supersymmetric particle,} {\sl Phys. Lett. B}
\textbf{444}, 367 (1998) [\hepph{9810360}].



\bibitem{ellis1} J. Ellis, T. Falk, K. A. Olive and M. Srednicki,
{\it Calculations of Neutralino-Stau Coannihilation Channels and
the Cosmologically Relevant Region of MSSM Parameter Space}, {\sl
Astropart. Phys. } \textbf{13}, 181 (2000)
[\texttt{\hepph{9905481}}].

\bibitem{cd1} M.E. G\'omez, G. Lazarides and C. Pallis, {\it Supersymmetric cold
dark matter with Yukawa unification}, {\sl Phys. Rev. D} {\bf 61},
123512 (2000) [\hepph{9907261}].



\bibitem{cd2} M.E.~G\'omez, G.~Lazarides and C.~Pallis,
{\it Yukawa unification, b --{\ensuremath{>}} s gamma and
Bino-Stau coannihilation,} {\it Phys. Lett. B} \textbf{487}, 313
(2000) [\hepph{0004028}].

\bibitem{cd3} S.~Khalil, G.~Lazarides and C.~Pallis,
{\it Cold dark matter and b --{\ensuremath{>}} s gamma in the Ho\v
rava-Witten theory,} {\sl Phys. Lett. B} \textbf{508}, 327 (2001)
[\hepph{0005021}].


\bibitem{phd} C.~Pallis, {\it Phenomenology and Cosmology of Supersymmetric
Grand Unified Theories,} \hepph{0007114}.


\bibitem{leo} I.~Antoniadis and G.K.~Leontaris, {\it A supersymmetric $SU(4) \times O(4)$ model}, {\sl Phys.  Lett. B }{\bf 216}, 333 (1989).


\bibitem{quasi} M.E. G\'omez, G. Lazarides and C. Pallis, {\it Yukawa
quasi-unification}, {\sl Nucl. Phys. } {\bf B638}, 165 (2002)
[\hepph{0203131}].

\bibitem{quasi1} M.E. G\'omez, G. Lazarides and C. Pallis, {\it On Yukawa
quasiunification with mu less than 0}, {\sl Phys. Rev. D} {\bf
67}, 097701 (2003) [\hepph{0301064}].


\bibitem{nick1} N. Karagiannakis, G. Lazarides and C. Pallis, {\it CMSSM with Yukawa
Quasi-Unification Revisited}, {\sl Phys. Lett. B} {\bf 704}, 43
(2011) [\arxiv{1107.0667}].

\bibitem{nick2} N. Karagiannakis, G. Lazarides and C. Pallis, {\it Constrained
Minimal Supersymmetric Standard Model with Generalized Yukawa
Quasi-Unification}, {\sl Phys. Rev. D} {\bf 87}, 055001 (2013)
[\arxiv{1212.0517}].

\bibitem{nick3} N. Karagiannakis, G. Lazarides and C. Pallis, {\it Probing the
hyperbolic branch/focus point region of the constrained minimal
supersymmetric standard model with generalized Yukawa
quasiunification}, {\sl Phys. Rev. D} {\bf 92}, no. 8, 085018
(2015) [\arxiv{1503.06186}].

\bibitem{nick4} N.~Karagiannakis, G.~Lazarides and C.~Pallis,
{\it Cold Dark Matter and Higgs Mass in the Constrained Minimal
Supersymmetric Standard Model with Generalized Yukawa
Quasi-Unification,} {\sl Int. J. Mod. Phys. A }\textbf{28},
1330048 (2013) [\arxiv{1305.2574}].

\bibitem{phdn}  N.~Karagiannakis,
\lt{\it Φαινομενολογία και Κοσμολογία των Υπερσυμμετρικών
Θεωριών}, 2015.

\bibitem{sh} R. Jeannerot  {\it et al.}, {\it Inflation and monopoles in
supersymmetric $SU(4)C \times SU(2)(L) \times SU(2)(R)$},
\jhep{10}{2000}{012} [{\tt hep-ph/0002151}].

\bibitem{nsh} R. Jeannerot, S. Khalil and G. Lazarides, {\it New shifted hybrid
inflation}, {\sl J. High Energy Phys.} {\bf 07}, 069 (2002)
[\hepph{0207244}].


\bibitem{sm} G. Lazarides and C. Panagiotakopoulos, {\it Smooth hybrid
inflation}, \prd{52}{1995}{559} [{\tt hep-ph/9506325}].

\bibitem{sm1} R.~Jeannerot, S.~Khalil and G.~Lazarides, {\it Leptogenesis
in smooth hybrid inflation}, \plb{506}{2001}{344} [{\tt
hep-ph/0103229}].

\bibitem{nsm} G. Lazarides and A. Vamvasakis, {\it New smooth hybrid inflation},
{\sl Phys. Rev. D} {\bf 76}, 083507 (2007) [\arxiv{0705.3786}].


\bibitem{ssh} G. Lazarides, I.N.R. Peddie and A. Vamvasakis, {\it Semi-shifted
hybrid inflation with $B-L$ cosmic strings}, {\sl Phys. Rev. D}
{\bf 78}, 043518 (2008) [\arxiv{0804.3661}].

\bibitem{stsm} G.~Lazarides and A.~Vamvasakis,
{\it Standard-smooth hybrid inflation,} {\sl Phys. Rev. D}
\textbf{76}, 123514 (2007) [\arxiv{0709.3362}].


\bibitem{fhi} G.R. Dvali, Q. Shafi and R.K. Schaefer, {\it Large
scale structure and supersymmetric inflation without fine tuning},
\prl{73}{1994}{1886} [{\tt hep-ph/9406319}].


\bibitem{com} G. Lazarides and C. Pallis, {\it Reducing the spectal index in F-term
hybrid inflation through a complementary modular inflation}, {\sl
Phys. Lett. B} {\bf 651}, 216 (2007) [\hepph{0702260}].


\bibitem{phdax}  A.~Vamvasakis,
{\it Phenomenology and Cosmology of Supersymmetric Grand Unified
Theories,} \arxiv{0907.4549}.

\bibitem{stssh} G.~Lazarides and C.~Panagiotakopoulos,
{\it Gravitational Waves from Double Hybrid Inflation,} {\sl Phys.
Rev. D} \textbf{92}, no.12, 123502 (2015) [\arxiv{1505.04926}].


\bibitem{lrtfhi} R. Armillis, G. Lazarides and C. Pallis, {\it Inflation,
leptogenesis, and Yukawa quasiunification within a supersymmetric
left-right model}, {\sl Phys. Rev. D} {\bf 89}, no. 6, 065032
(2014) [\arxiv{1309.6986}].


\bibitem{lrthi} C.~Pallis, {\it T-model Higgs inflation and metastable cosmic strings,}
\jhep{01}{2025}{178} [\arxiv{2409.14338}].



\bibitem{magg} G. Lazarides, M. Magg and Q. Shafi, {\it Phase Transitions and
Magnetic Monopoles in $SO(10)$}, {\sl Phys. Lett. B }{\bf 97}, 87
(1980).

\bibitem{kibble} T.W.B.~Kibble, G.~Lazarides and Q.~Shafi, {\it Strings in $SO(10)$,}
{\sl Phys. Lett. B} \textbf{113}, 237 (1982).

\bibitem{kibbledw} T.W.B.~Kibble, G.~Lazarides and Q.~Shafi,
{\it Walls Bounded by Strings,} {\sl Phys. Rev. D }\textbf{26},
435 (1982).


\bibitem{nano} J.~Ellis \etal,
{\it What is the source of the PTA GW signal?} {\sl Phys. Rev. D}
\textbf{109}, no.~2, 023522 (2024) [\arxiv{2308.08546}].



\bibitem{so} G.~Lazarides and Q.~Shafi,
{\it Monopoles, Strings, and Necklaces in $SO(10)$ and $E_6$,}
{\sl J. High Energy Phys.} \textbf{10}, 193 (2019)
[\arxiv{1904.06880}].


\bibitem{sotw} G.~Lazarides, Q.~Shafi and A.~Tiwari, {\it Composite topological
structures in $SO(10)$,} {\sl J. High Energy Phys.} \textbf{05},
119 (2023) [\arxiv{2303.15159}].

\bibitem{asfhi} G. Lazarides and C.~Pallis,
\textit{Probing the Supersymmetry-Mass Scale With F-term Hybrid
Inflation}, {\sl Phys.\ Rev.\ D} {\bf 108}, no. 9, 095055 (2023)
[\arxiv{2309.04848}].

\bibitem{pqfhi} G. Lazarides and C. Pallis, {\it F-Term Hybrid Inflation Followed by a
Peccei-Quinn Phase Transition}, {\sl Phys. Rev. D} {\bf 82},
063535 (2010) [\arxiv{1007.1558}].

\bibitem{r2} C.~Pallis,
{\it Linking Starobinsky-Type Inflation in no-Scale Supergravity
to MSSM,}  \jcap{04}{2014}{024}; Erratum: \jcap{07}{2017}{E01}
[\arxiv{1312.3623}].

\bibitem{hi}  G. Lazarides and C. Pallis, {\it Shift Symmetry and Higgs Inflation in
Supergravity with Observable Gravitational Waves}, {\sl J. High
Energy Phys.} {\bf 11}, 114 (2015) [\arxiv{1508.06682}].

\bibitem{qaisar}  Q.~Shafi, {\it Remembering George Lazaridis,}
{\sl PoS CORFU} \textbf{2023}, 309 (2024).

\end{thebibliography}
\end{document}